\documentclass[11pt]{article}
\usepackage{amssymb}
\usepackage{amsmath}
\usepackage{amsthm}
\usepackage{epsfig}

\newtheorem{thm}{Theorem}

\newtheorem{prop}[thm]{Proposition}

\newcommand{\R}{{\mathbb R}}

\newcommand{\di}{\displaystyle}

\title{Combining Approximation Algorithms for the Prize-Collecting TSP}

\author{Michel X. Goemans\thanks{MIT Department of
Mathematics. \texttt{goemans@math.mit.edu}.  Supported by NSF contract
CCF-0829878 and by ONR grant N00014-05-1-0148.}} 

\date{}

\begin{document}
\maketitle
\begin{abstract}
We present a $1.91457$-approximation algorithm for the
prize-collecting travelling salesman problem. This is obtained by
combining a randomized variant of a rounding algorithm of Bienstock et
al.~\cite{Bienstock93} and a primal-dual algorithm of Goemans and
Williamson \cite{GoemansW95}. 
\end{abstract}

\section{Introduction}
In the prize-collecting travelling salesman problem (PC-TSP), we are
given a vertex set $V$ (with $|V|=n$), a metric $c$ on $V\times V$
(i.e.\ $c$ satisfies (i) $c_{ij}=c_{ji}\geq 0$ for all $i,j\in V$ and
(ii) triangle inequality: $c_{ij}+c_{jk}\geq c_{ik}$ for all $i,j,k\in
V$), a special vertex $r\in V$ (the depot), penalties $\pi:
V\rightarrow \R_+$, and the goal is to find a cycle $T$ with $r\in
V(T)$ such that $$c(T)+\pi(V\setminus V(T))$$ is minimized, where
$c(T)=\sum_{(i,j)\in T} c_{ij}$, $\pi(S)=\sum_{i\in S} \pi_i$, and
$V(T)$ denotes the vertices spanned by $T$. 

The first constant approximation algorithm for PC-TSP was given by
Bienstock et al.~\cite{Bienstock93}. It is based on rouding the
optimum solution to a natural LP relaxation for the problem, and
provides a performance guarantee of $2.5$.  Goemans and Williamson
\cite{GoemansW95} have designed a primal-dual algorithm based on the same
LP relaxation, and this gives a 2-approximation algorithm for the
problem. In 1998, Goemans \cite{Goemans98} has shown that a simple
improvement of the algorithm of Bienstock et al.~gives a guarantee of
$2.055\cdots =\frac{1}{1-e^{-2/3}}$. Recently, Archer et
al.~\cite{Archer09} are the first to break the barrier of 2 and
provide an improvement of the primal-dual algorithm of Goemans and
Williamson; their performance guarantee is $1.990283$. In this note,
we show that by combining the rounding algorithm of Bienstock et al.~
and the primal-dual algorithm of Goemans and Williamson, we can obtain
a guarantee of $1.91456\cdots =\frac{1}{1-\frac{2}{3}e^{-1/3}}$. The analysis uses the technique in
\cite{Goemans98} together with an improved analysis of the primal-dual
algorithm as observed in \cite{ChudakRW04} and used in Archer et
al.~\cite{Archer09}. 

\section{Combining Approximation Algorithms}
 
We start by briefly reviewing the rounding result of Bienstock et
al.~\cite{Bienstock93}. Consider a classical LP relaxation of PC-TSP:
$$\begin{array}{llllll}
 & & & \mbox{Min} & \di\sum_{e\in E} c_e x_e
+ \sum_v \pi(v) (1-y_v) \\ 
& \lefteqn{\mbox{subject to:}} \\
 & & & & \di x(\delta(v))= 2y_v & v\in V\setminus \{r\} \\ 
(LP) & & & & x(\delta(S)) \geq 2 y_v  & S\subset V, r\notin S, v\in S\\ 
 & & & & 0\leq x_e \leq 1 & e\in E\\
 & & & & 0\leq y_v \leq 1 & v\in V \\
 & & & & y_r=1,
\end{array}$$
where $E$ denotes the edge set of the complete graph on $E$.  
For conciseness, we use $c(x)+\pi(\mathbf{1}-y)$ to denote the
objective function of this LP. Let $x^*, y^*$ be an optimum
solution of this LP relaxation, and let
$LP=c(x^*)+\pi(\mathbf{1}-y^*)$ denote its value. Bienstock et
al.~\cite{Bienstock93} show the following (based on the analysis of
Christofides' algorithm due to Wolsey \cite{Wolsey80} and Shmoys and
Williamson \cite{ShmoysW90}).
 
\begin{prop}[Bienstock et al.] \label{prop:r}
Let $0<\gamma\leq 1$ and let $S(\gamma)=\{v: y^*_v\geq
\gamma\}$. Let $T_\gamma$ denote the cycle on $S(\gamma)$ output by
Christofides' algorithm when given $S(\gamma)$ as vertex set. Then:
$$c(T_\gamma) \leq \frac{3}{2\gamma} c(x^*).$$
\end{prop}

The 2.5-approximation algorithm can then be derived by setting
$\gamma=\frac{3}{5}$ since we get $c(T_{3/5})\leq \frac{5}{2} c(x^*)$
and $\pi(V\setminus S(3/5))\leq \frac{5}{2} \pi(\mathbf{1}-y^*)$. In
\cite{Goemans98}, we have shown that one can get a
better performance guarantee by taking the best cycle output over all
possible values of $\gamma$; notice that this leads to at most $n-1$
different cycles. 

The primal-dual algorithm in \cite{GoemansW95} constructs a cycle $T$
and a dual solution to the linear programming relaxation above such
that their values are within a factor 2 of each other, showing a
performance guarantee of 2 since the value of any dual solution is a
lower bound on $LP$. Chudak, Roughgarden and Williamson
\cite{ChudakRW04} (see their Theorem 2.1) observe that the analysis of
\cite{GoemansW95} actually shows a stronger guarantee on the penalty
side of the objective function, namely that the cycle $T$ returned
satisfies:
\begin{equation} \label{eq1} 
c(T) + \left(2-\frac{1}{n-1}\right) \pi(V\setminus V(T)) \leq
\left(2-\frac{1}{n-1}\right) LP.
\end{equation} 
This increased factor on the penalty side is exploited in Archer et
al.~\cite{Archer09}, and this motivated the result in this
note. Suppose now that we apply the primal-dual algorithm to an
instance in which we replace the penalties $\pi(\cdot)$ by
$\pi'(\cdot)$ given by
\begin{equation} \label{eqpi}
\pi'(v) = \frac{1}{2-1/(n-1)} \pi(v).
\end{equation}
Thus, (\ref{eq1}) implies
that the cycle $T$ returned satisfies:
\begin{equation} \label{eq2}
c(T) + \pi(V\setminus V(T))\leq \left(2-\frac{1}{n-1}\right) LP',
\end{equation}
where $LP'$ denotes the LP value for the penalties $\pi'(\cdot)$. As
the optimum solution $x^*$, $y^*$ of LP (with penalties $\pi(\cdot)$)
is feasible for the linear programming relaxation with penalties
$\pi'(\cdot)$, we derive that the cycle $T_{pd}$ output satisfies:
\begin{eqnarray*}
c(T_{pd}) + \pi(V\setminus V(T_{pd})) & = & c(T_{pd})+
\left(2-\frac{1}{n-1}\right) \pi'(V\setminus V(T_{pd})) \\
& \leq & \left(2-\frac{1}{n-1}\right) LP' \\
& \leq & \left(2-\frac{1}{n-1}\right) \left( c(x^*) +
\pi'(\mathbf{1}-y^*)\right) \\
& = &  \left(2-\frac{1}{n-1}\right) c(x^*) +
\pi(\mathbf{1}-y^*).
\end{eqnarray*}
Summarizing:
\begin{prop} \label{prop:pd}
The primal-dual algorithm applied to an instance with penalties
$\pi'(\cdot)$ given by (\ref{eqpi}) outputs a cycle $T_{pd}$ such that 
$$c(T_{pd}) + \pi(V\setminus V(T_{pd})) \leq 2 c(x^*) + \pi(\mathbf{1}-y^*).$$
\end{prop}

We claim that the best of the algorithms given in Propositions
\ref{prop:r} and \ref{prop:pd} gives a better than 2 approximation
guarantee for PC-TSP. 

\begin{thm} \label{thm3}
Let $$H=\min(\min_\gamma (c(T_\gamma)+\pi(V\setminus
V(\gamma))),c(T_{pd})+\pi(V\setminus V(T_{pd}))).$$ Then 
$$H\leq \alpha \left(c(x^*) + \pi(\mathbf{1}-y^*)\right)=\alpha LP,$$
where $\alpha =\frac{1}{1-\frac{2}{3}e^{-1/3}}<1.91457$.
\end{thm}
As mentioned earlier, the minimum in the theorem involves only $n$
different algorithms as we need only to consider values $\gamma$ equal
to some $y^*_v$.
 
\begin{proof}
We construct an appropriate probability distribution over all the
algorithms involved such that the expected cost of the solution
produced is at most $\alpha \left(c(x^*) + \pi(\mathbf{1}-y^*)\right)$. 

First, assume that we select $\gamma$ randomly (according to a certain
distribution to be specified). Then, by Proposition \ref{prop:r}, we
have that $$E[c(T_\gamma)]\leq \frac{3}{2}
E\left[\frac{1}{\gamma}\right]c(x^*),$$ while the expected penalty we
have to pay is 
$$E[\pi(V\setminus V(\gamma))]=\sum_{v\in V} Pr[\gamma >y^*(v)]
\pi(v).$$
Thus, the overall expected cost is:
\begin{equation} \label{eqa}
E[c(T_\gamma)+\pi(V\setminus V(\gamma))]\leq \frac{3}{2}
E\left[\frac{1}{\gamma}\right]c(x^*) + \sum_{v\in V} Pr[\gamma >y^*(v)]
\pi(v).
\end{equation}
Assume now that $\gamma$ is chosen uniformly between
$a=e^{-1/3}=0.71653\cdots$ and $1$. Then,
$$E\left[\frac{1}{\gamma}\right]=\int_a^1 \frac{1}{1-a}\frac{1}{x} dx =
-\frac{\ln(a)}{1-a}=\frac{1}{3(1-a)}=\frac{1}{3(1-e^{-1/3})},$$
and 
$$ Pr[\gamma >y]=\left\{\begin{array}{ll} \frac{1-y}{1-a} & a\leq y
\leq 1 \\ 1 \leq \frac{1-y}{1-a} & 0\leq y \leq a. \end{array}\right.
$$ Therefore, (\ref{eqa}) becomes:
\begin{equation} \label{eqb}
E[c(T_\gamma)+\pi(V\setminus V(\gamma))]\leq \frac{1}{2(1-e^{-1/3})}
c(x^*) + \frac{1}{1-e^{-1/3}} \pi(\mathbf{1}-y^*).
\end{equation}

Suppose we now select, with probability $p$, the primal-dual algorithm
as given in Proposition \ref{prop:pd} or, with probability $1-p$, the
rounding algorithm with $\gamma$ chosen randomly acording to
$\gamma\sim U[e^{-1/3},1]$. From (\ref{eqb}) and Proposition
\ref{prop:pd}, we get that the expected cost $E^*$ of the resulting
algorithm satisfies:
$$E^*\leq \left(2p+(1-p) \frac{1}{2(1-e^{-1/3})}\right)c(x^*) +
\left(p+(1-p)\frac{1}{1-e^{-1/3}}\right)  \pi(\mathbf{1}-y^*).$$
Choosing $p=(1-p)\frac{1}{2(1-e^{-1/3})}$,
i.e. $p=\frac{1}{3-2e^{-1/3}}$, we get 
$$E^*\leq 3p(c(x^*)+\pi(\mathbf{1}-y^*))= 3p LP.$$
Therefore, the best of the algorithms involved outputs a solution of
cost at most $3p LP=\alpha LP$ where $$\alpha
=\frac{1}{1-\frac{2}{3}e^{-1/3}}<1.91457.$$ 
\end{proof}

One can show that the probability distribution given in the proof is
optimal for the purpose of this proof; this is left as an exercise for
the reader. 

Theorem \ref{thm3} shows that the linear programming relaxation of
PC-TSP has an integrality gap bounded by $1.91457$; in contrast, the
result of Archer et al.~\cite{Archer09} does not imply a better than 2
bound on the integrality gap.

As a final remark, if we replace Christofides' algorithm with an algorithm
for the symmetric TSP that outputs a solution within a factor $\beta$
of the standard LP relaxation for the TSP then the approach described
in this note gives a guarantee of 
$$\frac{1}{1-\frac{1}{\beta}e^{1-2/\beta}}$$ for PC-TSP.

\end{document}